\documentclass[twocolumn,aps,prb,superscriptaddress,longbibliography,floatfix]{revtex4-2}

\usepackage{xr-hyper}
\usepackage{times}
\usepackage{graphicx}
\usepackage{amsfonts}
\usepackage{amsmath, amsthm, amssymb, mathrsfs}
\usepackage{empheq}
\usepackage{dsfont}
\usepackage{xcolor}
\usepackage{cancel}
\usepackage{scalerel}
\usepackage{standalone}
\usepackage{stmaryrd}
\usepackage{microtype}
\usepackage{tikz}
\usetikzlibrary{decorations.markings}
\usetikzlibrary{shapes,positioning,arrows}
\usepackage{bm}
\usepackage{siunitx}
\usepackage{slashed}
\usepackage{siunitx}
\usepackage[colorlinks=true, allcolors=black, citecolor=blue, linkcolor=blue, urlcolor=blue]{hyperref}
\usepackage[capitalize]{cleveref}

\allowdisplaybreaks

\usepackage{graphicx}
\usepackage[caption=false]{subfig}
\graphicspath{{figures/}}

\renewcommand{\l}{\left}
\renewcommand{\r}{\right}

\newcommand{\s}{\sigma}

\newcommand{\ie}{\emph{i.e.}~}
\newcommand{\eg}{\emph{e.g.}~}

\newcommand{\be}{\begin{equation}}
\newcommand{\ee}{\end{equation}}
\newcommand{\bse}{\begin{subequations}}
\newcommand{\ese}{\end{subequations}}
\newcommand{\bal}{\begin{align}}
\newcommand{\eal}{\end{align}}

\newcommand{\up}{\uparrow}
\newcommand{\down}{\downarrow}

\newcommand{\oMn}{\omega_n}
\renewcommand{\l}{\left}
\renewcommand{\r}{\right}

\renewcommand{\c}[2]{c_{\bm{k}#1}^{\scaleto{(#2)}{6pt}}}
\newcommand{\cd}[2]{c_{\bm{k}#1}^{\scaleto{(#2)}{6pt}\dagger}}
\newcommand{\cm}[2]{c_{-\bm{k}#1}^{\scaleto{(#2)}{6pt}}}
\newcommand{\cdm}[2]{c_{-\bm{k}#1}^{\scaleto{(#2)}{6pt}\dagger}}
\newcommand{\x}[1]{\xi_{\bm{k}}^{\scaleto{(#1)}{6pt}}}
\newcommand{\tm}[1]{\hat{\tau}_{#1}}

\usepackage{soul}

\usepackage[integrals]{wasysym}

\begin{document}

\title{High magnetic field superconductivity in a two-band superconductor}

\author{Tancredi Salamone}
\affiliation{Center for Quantum Spintronics, Department of Physics, NTNU, Norwegian University of Science and Technology, NO-7491 Trondheim, Norway}
\author{Henning G. Hugdal}
\affiliation{Center for Quantum Spintronics, Department of Physics, NTNU, Norwegian University of Science and Technology, NO-7491 Trondheim, Norway}
\author{Sol H. Jacobsen}
\affiliation{Center for Quantum Spintronics, Department of Physics, NTNU, Norwegian University of Science and Technology, NO-7491 Trondheim, Norway}
\author{Morten Amundsen}
\affiliation{Nordita, KTH Royal Institute of Technology and Stockholm University, Hannes Alfvéns väg 12, SE-106 91 Stockholm, Sweden}

\date{\today}

\begin{abstract}
When applying an external magnetic field to a superconductor, orbital and Pauli paramagnetic pairbreaking effects govern the limit of the upper critical magnetic field that can be supported before superconductivity breaks down. Experimental studies have shown that many multiband superconductors exhibit values of the upper critical magnetic field that violate the theoretically predicted limit, giving rise to many studies treating the underlying mechanisms that allow this. In this work we consider spin-splitting induced by an external magnetic field in a superconductor with two relevant bands close to the Fermi level, and show that the presence of interband superconducting pairing produces high-field reentrant superconductivity violating the Pauli-Chandrasekhar-Clogston limit for the value of the upper critical magnetic field.
\end{abstract}

\maketitle

\section{Introduction}

Over the past two decades, multiband superconductors have been attracting great interest because of increasing experimental evidence of interesting effects not achievable in single band systems. For instance, in the first-ever discovered multiband superconductor, MgB\textsubscript{2}, Leggett modes have been observed \cite{Blumberg2007} and more recently optically-controlled \cite{Giorgianni2019}. Moreover, spontaneous time reversal symmetry breaking has been reported in Ba\textsubscript{$1\!-\!x$}K\textsubscript{$x$}Fe\textsubscript{2}As\textsubscript{2}, Sr\textsubscript{2}RuO\textsubscript{4}, UPt\textsubscript{3} and many other multiband systems (for an exhaustive review see \cite{Ghosh_2020}). 

Since the extension of the Bardeen–Cooper–Schrieffer (BCS) theory to multiband systems by Suhl, Matthias and Walker \cite{Suhl1959} and Moskalenko \cite{Moskalenko1959}, many studies have focused on a theoretical understanding of the effects of a multiband description of superconductors \cite{Korchorbe1993,Tahir-Kheli1998,Caprara2013,Marciani2013,Komendova2015,Cea2016,Komendova2017,Vargas-Paredes2020,Barzykin2009,Iskin2005,Iskin2006,Yerin2019}. In general, if two bands are close to each other or hybridized, it is possible to form interband Cooper pairs (see \eg \cite{Schaffer2013}), where the electrons comprising the pairs come from two distinct bands. 
Research on interband pairing has been rather limited, but studies have found that it affects Josephson tunneling \cite{Tahir-Kheli1998}, it is an important factor in obtaining an anomalous Hall effect \cite{Taylor2012}, it can produce gapless states \cite{Vargas-Paredes2020}, and it influences the BCS-BEC crossover \cite{Bremm2021}. We note that, in the context of this work, the term interband will always refer to Cooper pairs formed by electrons in distinct bands, and it must not be confused with the same term often found in the literature, also called pair-hopping, referring to the hopping of intraband pairs, \ie formed by electrons in the same band, between different bands.

Among multiband superconductors, MgB\textsubscript{2} and Fe-Based Superconductors (FeBS) are particularly interesting because of their high critical temperatures and upper critical magnetic fields. For instance, the critical temperature is $39\si{K}$ for MgB\textsubscript{2} \cite{Nagamatsu2001}, $55\si{K}$ for SmO\textsubscript{$1\!-\!x$}F\textsubscript{$x$}FeAs \cite{Zhi_An_2008} and $65\si{K}$ in FeSe films on SrTiO\textsubscript{3} substrate \cite{Wang_2012}, whereas the zero temperature estimated values of the upper critical field are $25\si{T}$ in single crystal MgB\textsubscript{2} \cite{Xu2001}, $70\si{T}$ in C-doped MgB\textsubscript{2} thin films \cite{Braccini2005} and up to $300\si{T}$ in FeBS \cite{Sefat2008,Chen2008,Ying2008,Altarawneh2008,Yuan2009,Smylie2019}. It is worth noting, however, that these values are often extrapolated from low-magnetic field data obtained close to the critical temperature. Therefore, the extrapolated low-temperature dependence of the upper critical field and its $T=0$ value may be a bad estimate. However, the application of pulsed fields allows to reach higher magnetic field values and obtain more reliable estimations \cite{Yuan2009}.

The simultaneous presence of superconductivity and magnetism is of great interest for the field of superconducting spintronics \cite{Linder2015} where the proximity effect is exploited to achieve dissipationless information transport. However, these two phenomena are often mutually exclusive since two effects contribute to destroying superconductivity when an external magnetic field is applied: the orbital and Pauli paramagnetic pairbreaking effects \cite{Clogston1962,Chandrasekhar1962}. The orbital effect describes the breaking of Cooper pairs when the kinetic energy of electrons, resulting from the momentum acquired in a magnetic field, exceeds the superconducting gap. On the other hand, paramagnetic pairbreaking occurs when Cooper pairing becomes energetically unfavourable as the Zeeman energy of the electrons overcomes the superconducting gap. This happens when the exchange energy reaches a value given by the Pauli, or equivalently the Chandrasekhar-Clogston, limit $h_c=\Delta_0/\sqrt{2}=1.86T_c$, 
where $\Delta_0$ is the value of the superconducting gap at zero temperature and zero applied field, and $T_c$ is the superconducting critical temperature.

\begin{figure*}
    \centering
    \includegraphics[width=1.75\columnwidth]{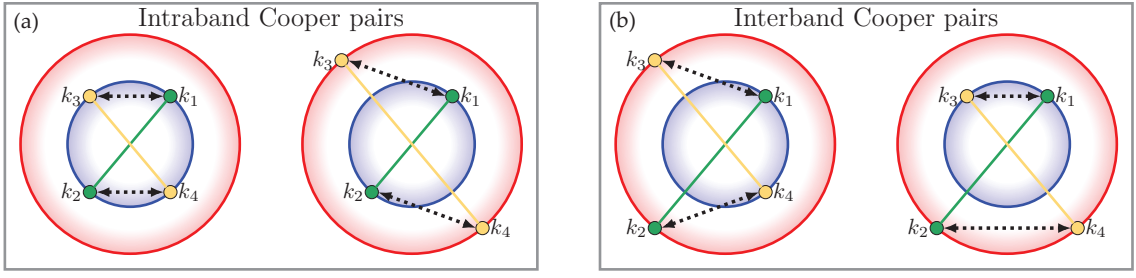}
    \caption{Illustration of some of the superconducting pairing processes possible in a two-band system for (a) intraband and (b) interband Cooper pairs. The red and blue circles represent the Fermi surfaces of the two bands, the green and yellow lines identify the electrons forming the Cooper pairs and the dotted lines indicates the scattering processes.}
    \label{fig:pairhop}
\end{figure*}

Various mechanisms producing the limit violation, or enhancement of upper critical fields have been proposed, \eg scattering by non-magnetic impurities \cite{Gurevich2003,Gurevich2007}, spin-triplet superconductivity \cite{Ran2019}, Fulde-Ferrell-Larkin-Ovchinnikov (FFLO) pairing \cite{Ptok2013,Ptok2014}, strong superconducting coupling, spin-orbit coupling \cite{Xi2016,Nam2016}, application of a voltage bias \cite{Bobkova2011,Ouassou2018}, the proximity of two bands to each other \cite{Shulga1998,Ghanbari2022,Horhold2022}, and pair-hopping in three band superconductors \cite{Yerin2013,Marques2015}. Furthermore, many experimental works have reported evidence for upper critical magnetic field violating the above limits in multiple superconductors, \eg NbSe\textsubscript{2} \cite{Xi2016,Kuzmanovic2022}, iron pnictides \cite{Ghannadzadeh2014,Xing2017}, lanthanide infinite-layer nickelate \cite{Chow2022}, moiré graphene \cite{Cao2021}, organic superconductors \cite{Balicas2001,Uji2001}, Eu-Sn molybdenum chalcogenide \cite{Meul1984}, URhGe \cite{Levy2005}. Refs\cite{Cao2021,Balicas2001,Uji2001,Meul1984,Levy2005} are particularly interesting for the purpose of this work because their results present two disconnected superconducting domains, for small and large external magnetic field, due to non-spin-singlet Cooper pairs in Ref.\cite{Cao2021} and to the Jaccarino-Peter effect in Refs.\cite{Balicas2001,Uji2001,Meul1984}.

In this work, we present a simple mechanism which allows to overcome the conventional limit. We consider a two band \textit{s}-wave superconductor in the presence of an external magnetic field and demonstrate that the inclusion of interband pairing allows to overcome the limiting value of the critical field found in conventional superconductors. Within a BCS framework we show that for a certain range of parameters, our system exhibits superconductivity for significantly high values of the external magnetic field. Furthermore, we show that the system can exhibit two separate superconducting domains, for small and large external magnetic field, and we provide an explanation of the mechanism producing these results.

\section{Theory}

We consider the following mean-field Hamiltonian for a two bands spin-split superconductor with both intra- and interband spin-singlet superconducting coupling:

\be\begin{split}
	\mathcal{H}=&\sum_{\bm{k}\s}(\x{1}\!-\!E_c\!-\!\s h)\cd{\s}{1}\!\c{\s}{1}\!+\!\sum_{\bm{k}\s}(\x{2}\!+\!E_c\!-\!\s h)\cd{\s}{2}\!\c{\s}{2}\\
	&-\sum_{\bm{k}}\sum_{\alpha,\alpha'=1,2}\l(\Delta_{\alpha\alpha'}(\bm{k})\cd{\up}{\alpha}\cdm{\down}{\alpha'}+\mathrm{h.c.}\r)\\
	\label{eq:hamiltonian}
\end{split}\ee

\noindent where the operator $\cd{\s}{\alpha}$ ($\c{\s}{\alpha}$) creates (destroys) an electron in band $\alpha$ with dispersion $\x{\alpha}=\varepsilon^{\alpha}_{\bm{k}}-\mu$ and spin $\s$, $E_c$ is half the band separation and $h$ is the externally applied in-plane magnetic field. The superconducting order parameters $\Delta_{\alpha\alpha'}(\bm{k})$ are defined by:

\be
    \Delta_{\alpha\alpha'}(\bm{k})=\frac{T}{V}\sum_{\oMn}\sum_{\beta,\beta'=1,2}\sum_{\bm{k}'}g_{\alpha\alpha',\beta\beta'}(\bm{k},\bm{k}')F^{\beta\beta'}(\bm{k}',\oMn),
    \label{eq:gapEq}
\ee

\noindent where $\oMn=(2n+1)\pi/\beta$ is the fermionic Matsubara frequency and $F^{\beta\beta'}$ is the anomalous component of the Green's function. 

The superconducting coupling matrix $g_{\alpha\alpha',\beta\beta'}(\bm{k},\bm{k}')$ defines the different coupling processes, the terms $g_{\alpha\alpha,\alpha'\alpha'}$ describe processes involving intraband Cooper pairs: formed by electrons in the same band, hopping between the same band ($\alpha=\alpha'$) or different bands ($\alpha\neq\alpha'$). This last term is often referred to in the literature as interband scattering, or pair hopping, and must not be confused with the use we make of the term interband. Processes involving interband pairs instead are described by those elements $g_{\alpha\alpha',\beta\beta'}$ with $\alpha\neq\alpha'$ and/or $\beta\neq\beta'$. 
The superconducting pairing processes relevant for the purpose of this work are illustrated in \cref{fig:pairhop} for a two-band system.

The terms $\Delta_{11}$ and $\Delta_{22}$ represent the intraband order parameters, while $\Delta_{12}=\Delta_{21}$ is the interband order parameter.

In the basis defined by $\hat{\psi}_{\bm{k}}^\dagger=\l(\cd{\up}{1},\cm{\down}{1},\cd{\up}{2},\cm{\down}{2}\r)$ the inverse Green's function for the Hamiltonian of \cref{eq:hamiltonian} is:

\begin{widetext}
\be
\bm{\mathcal{G}}_0^{-1}=\begin{pmatrix} (i\oMn+h)\tm{0}-\tilde{\xi}_1\tm{3}+\Delta_{11}\tm{1} & \Delta_{12}\tm{1} \\ \Delta_{12}\tm{1} & (i\oMn+h)\tm{0}-\tilde{\xi}_2\tm{3}+\Delta_{22}\tm{1} \end{pmatrix}.
\label{eq:invG}
\ee
\end{widetext}

\noindent where $\tilde{\xi}_1=\x{1}-E_c$, $\tilde{\xi}_2=\x{2}+E_c$ and we restricted to s-wave pairing (real order parameter). The two branches of the BCS quasiparticle excitation spectrum, $E_+$ and $E_-$, are obtained by defining $\mathrm{det}\bm{\mathcal{G}}_0^{-1}\equiv\l(\tilde{\omega}_n^2+E_+^2\r)\l(\tilde{\omega}_n^2+E_-^2\r)$, where $\tilde{\omega}_n=\oMn-ih$. Inverting \cref{eq:invG} we obtain the Green's function of the system which allows us to analyze when superconductivity occurs. Details of the form of the Green’s function are given in \cref{appendix:Green}.

\begin{figure}
    \centering
    \includegraphics[width=\columnwidth]{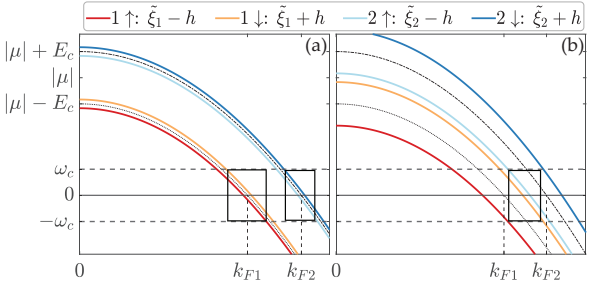}
    \caption{Band structure of the system in the presence of an exchange field, for (a) small and (b) large values of the field and generic values of $E_c$, $\gamma$, $\omega_c$, $\mu$. The rectangles indicate the regions where Cooper pair formation can occur.}
    \label{fig:bands}
\end{figure}

\section{Results}

We consider a two bands system with spin-split hole-like parabolic bands: $\x{\alpha}=-k^2/(2m_\alpha)-\mu$ with $m_\alpha$ the effective mass of the band $\alpha$ and $\mu$ the chemical potential. Defining the ratio between the two band masses as $\gamma=m_1/m_2$ we can write $\x{2}=\gamma\x{1}+(\gamma-1)\mu$. The band structure of the system for generic values of $E_c$, $\gamma$, $\mu$ and two different limits of $h$ is represented in \cref{fig:bands}, where we also show the energy cutoff of the effective attractive interaction $\omega_c$. By examining the case of small spin-splitting represented in \cref{fig:bands}(a), we note that intraband pairing is favored, since bands with the same band index and opposite spin are close to each other. On the other hand, the presence of a large spin-splitting field can bring two bands with different band and spin indices closer to each other, favoring an interband pairing mechanism, as shown in \cref{fig:bands}(b).

To simplify the problem, we neglect those scattering processes connecting interband to intraband Cooper pairs and vice versa, \ie we set  $g_{\alpha\alpha,\alpha\beta}=g_{\alpha\beta,\alpha\alpha}=0$ with $\alpha\neq\beta$. This simplification is justified by the fact that, for sufficiently large $E_c$, either interband or intraband pairing processes will dominate over the other, depending on $h$ and $\gamma$. Simultaneously, the processes of intraband pairs hopping in interband pairs, or vice versa, are either forbidden due to energy or momentum conservation, or strongly suppressed compared to the non-mixing processes, since they would meet energy and momentum conservation criteria for significantly smaller ranges. 


With these simplifications, from \cref{eq:gapEq}, it is clear that the gap equation for the intraband and interband order parameters are decoupled and we can solve them separately. We set the chemical potential $\mu$ as the energy scale and choose the energy cutoff for the effective attractive interaction and the band separation to be $\omega_c=0.2\l|\mu\r|$ and $E_c=0.05\l|\mu\r|$, respectively. Furthermore, we consider dimensionless superconducting coupling constants: $\lambda^{\mathrm{intra}}_{\alpha\beta}=N_\alpha(0)g_{\alpha\alpha,\beta\beta}$ and $\lambda^{\mathrm{inter}}=N_1(0)g_{\alpha\beta,\alpha\beta},\,(\alpha\neq\beta)$, where $N_\alpha(0)$ is the density of states at the Fermi energy for the band $\alpha$. Their values are chosen to be $\lambda^{\mathrm{intra}}_{11}=0.3$, $\lambda^{\mathrm{inter}}=\lambda^{\mathrm{intra}}_{22}=2\lambda^{\mathrm{intra}}_{12}=0.2$. We assume that $\lambda^{\mathrm{inter}}$ can be taken to be larger than some $\lambda^{\mathrm{intra}}_{\alpha\beta}$ because of the external magnetic field, which may bring two different bands with opposite spin very close to each other in energy, see \eg \cref{fig:bands}(b). 


Having chosen the values of our parameters we can establish whether the system exhibits superconductivity. We determine the critical values of temperature $T_c$ and exchange field $h_c$ of our system for different values of the effective mass ratio $\gamma$ by linearizing \cref{eq:gapEq} with respect to $\Delta_{\alpha\alpha'}$, separately for interband ($\alpha\neq\alpha'$) and intraband ($\alpha=\alpha'$) superconducting pairing. The linearized gap equations are derived in \cref{appendix:inter_gapeq,appendix:intra_gapeq}. In the interband case we have the following equation

\be\begin{split}
    \frac{1}{\lambda^{\mathrm{inter}}}=\int_{-\omega_c}^{\omega_c}&\frac{d\xi}{2(\tilde{\xi}_1+\tilde{\xi}_2)}\sum_{\alpha=1,2}\biggl\{\tanh\l[\frac{\beta_c}{2}\l(\tilde{\xi}_\alpha+h_c\r)\r]\\
    &+\tanh\l[\frac{\beta_c}{2}\l(\tilde{\xi}_\alpha-h_c\r)\r]\biggr\},
    \label{eq:lingapinter}
\end{split}\ee

\noindent where $\beta_c=1/T_c$, $\tilde{\xi}_1=\xi-E_c$, and $\tilde{\xi}_2=\gamma\xi+(\gamma-1)\mu+E_c$. The interband critical parameters are the values satisfying \cref{eq:lingapinter}. We note that by setting $E_c=0$ and $\gamma=1$, we get $\tilde{\xi}_1=\tilde{\xi}_2=\xi$, and \cref{eq:lingapinter} reduces to that of a single band superconductor in external magnetic field, reported in \cref{appendix:inter_gapeq}. In the intraband case we have the following system of equations

\be
    \begin{pmatrix} \delta_1 \\ \delta_2 \end{pmatrix}\begin{pmatrix} \lambda_{11}I_1(T_c,h_c)-1 & \lambda_{12}I_2(T_c,h_c) \\ \lambda_{21}I_1(T_c,h_c) & \lambda_{22}I_2(T_c,h_c)-1 \end{pmatrix}=0,
    \label{eq:lingapintra}
\ee

\noindent where the terms $\delta_\alpha$ are defined through $\Delta_{\alpha\alpha}=\epsilon\delta_\alpha$, $\epsilon(T=T_c)=0$, and

\begin{equation}
    I_\alpha(T,h)\!=\!\!\int_{-\omega_c}^{\omega_c}\!\frac{d\xi}{4\tilde{\xi}_\alpha}\!\l\{\!\tanh\!\l[\!\frac{\beta}{2}\!\l(\!\tilde{\xi}_\alpha\!+\!h\r)\!\r]\!+\!\tanh\!\l[\!\frac{\beta}{2}\!\l(\!\tilde{\xi}_\alpha\!-\!h\!\r)\!\r]\!\r\}\!,
\end{equation}

\noindent with $\alpha=1,2$, and $\beta=1/T$. The intraband critical parameters are found by setting to zero the determinant of the matrix in \cref{eq:lingapintra}. More details on the procedure used to obtain the critical parameters are reported in \cref{appendix:hc(T)}.

The results are shown in \cref{fig:comb_hc_Ec0.05}, displaying reentrant superconductivity. The figure shows the inter- and intraband superconducting domains, delimited by the $h_c(T)$ curves, for different values of $\gamma$, the lines correspond to the critical values of temperature and exchange field, while the colored area identifies the superconducting region, with dark gray color identifying the interband regions and light gray the intraband. From \cref{fig:comb_hc_Ec0.05}, we immediately note a substantial difference between the superconductivity induced by each type of coupling. The intraband region develops around $h=0$ for all the values of $\gamma$, with the highest critical temperature obtained in absence of any external field. On the other hand, the interband regions show an interesting behavior: except for the case $\gamma=1.1$, they are not present in the zero-field case but around a certain finite value of $h$, thus their appearance is conditioned to the presence of an external field. Therefore, analyzing \cref{fig:comb_hc_Ec0.05}, we note that for certain values of $\gamma$ it is possible to find two distinct/disconnected superconducting regions: the intraband one for low exchange field values and the interband one developing around substantially higher values of the exchange field. We note that the value $\gamma=1.1$ as a certain significance, since it approximately corresponds to the point where the difference in band curvature compensate for the band separation $2E_c$, so that the Fermi momenta of the two bands become the same. Hence, any $h\neq0$ has a detrimental effect for superconductivity. 

\begin{figure}
    \centering
    \includegraphics[width=\columnwidth]{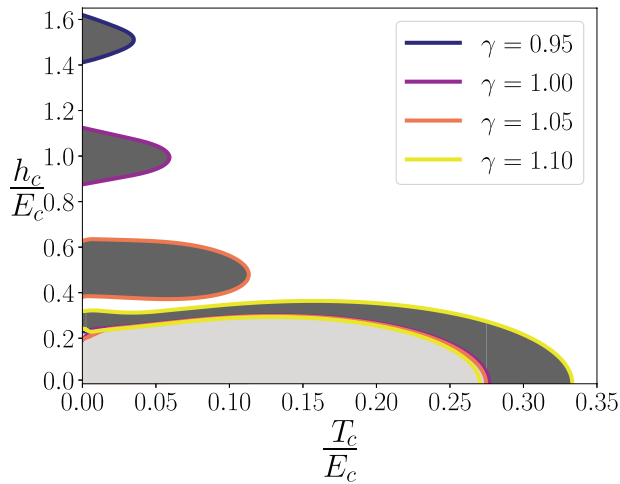}
    \caption{Critical field $h_c$ as a function of temperature $T$ for different values of the band mass ratio $\gamma$, for $E_c=0.05\l|\mu\r|$, $\lambda^{\mathrm{intra}}_{11}=0.3$, $\lambda^{\mathrm{inter}}=\lambda^{\mathrm{intra}}_{22}=2\lambda^{\mathrm{intra}}_{12}=0.2$, and $\omega_c=0.2\l|\mu\r|$. The dark (light) gray areas represent the region of interband (intraband) superconducting coupling. Except for the case $\gamma=1.1$, we note a reentrant superconductivity, with the two disconnected intra- and interband superconducting domains.}
    \label{fig:comb_hc_Ec0.05}
\end{figure}

These results can be explained by considering the band structure provided in \cref{fig:bands} and the conditions for superconducting pairing to occur. In the BCS theory, Cooper pairs are formed by electrons with opposite momenta close to the Fermi momentum, and energies in a ``shell" of width $\omega_c$ around the Fermi energy. Therefore, only electrons meeting these two criteria can form Cooper pairs. In a multiband system with spin-split bands, each band will have its own intervals of energies and momenta where this can be realized. When the intervals of two different bands overlap, Cooper pair formation among them is feasible. These overlap regions are represented by the rectangles in \cref{fig:bands}.

In the absence of spin-splitting, and in the case of intraband pairing Cooper pair formation is of course always possible. When an external field is applied, instead, the spin-up and spin-down bands separate from each other, with their distance increasing with $h$, thus reducing the size of the overlap region where the formation of Cooper pairs is possible. Consequently, for purely intraband pairing, the critical temperature of the system has a maximum at zero field.

This picture changes when considering the case of interband pairing. While in the intraband case the exchange field pulls the two pairs of spin-split bands apart from each other, in the interband case it can have the opposite effect. Superconductivity involving interband pairs can result from different cases. With spin-split bands, we can have formation of Cooper pairs with (i), one electron in band $1\down$ and the other in $2\up$ and (ii), one electron in band $1\up$ and the other in band $2\down$. Without spin-splitting instead we have (iii), one electron in band 1 and the other in band 2. The choice of the parameters $E_c$, $h$ and $\gamma$ determines the size of the overlap between the different bands and the size of the superconducting coupling constant $\lambda^{\mathrm{inter}}$ influences in which of the three cases superconductivity is realized.

Qualitatively, in each case the maximum in the size of the overlap, and thus in the superconducting critical temperature, occurs when the Fermi momenta of the two bands involved are equal to each other. Therefore, equating the Fermi momenta of the two bands for each of these cases, we can obtain, as a function of the other parameters, the value of $h$ which determines the match of the momenta, allowing for Cooper pair formation. In case (i) we have $k_{F1\down}=\sqrt{-2m(\mu+E_c-h)}$ and $k_{F2\up}=\sqrt{-2m_1/\gamma(\mu-E_c+h)}$, equating the two we obtain: 

\be
h=\frac{\gamma-1}{\gamma+1}\mu+E_c.
\label{eq:hpos1d2u}
\ee

\noindent In case (ii) we get $k_{F1\up}=\sqrt{-2m_1(\mu+E_c+h)}$ and $k_{F2\down}=\sqrt{-2m_1/\gamma(\mu-E_c-h)}$, yielding:

\be
h=-\frac{\gamma-1}{\gamma+1}\mu-E_c.
\label{eq:hpos1u2d}
\ee

\noindent In case (iii) it is clear that $h=0$.

We note that the value of $h_c$ at which the interband superconducting domains present a maximum in $T_c$, depends only on the band parameters, $\mu$, $E_c$ and $\gamma$, and not on the superconducting coupling constants. These instead, together with other factors, such as \eg impurity scattering, influence the value of $T_c$, and therefore, whether superconductivity is realized.

\begin{figure}
    \centering
    \includegraphics[width=\columnwidth]{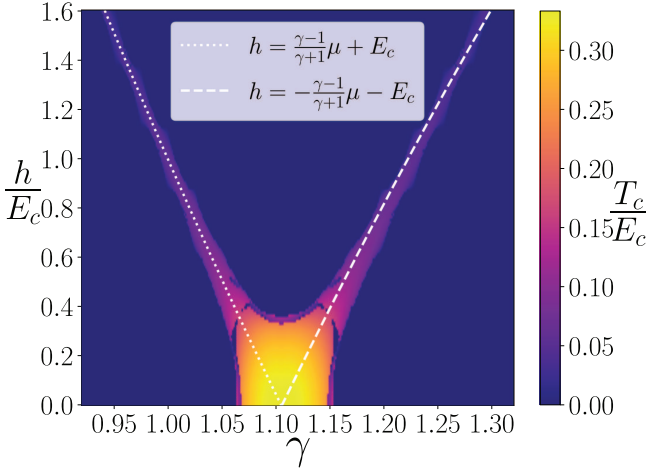}
    \caption{Critical temperature of the interband superconducting domains as a function of $\gamma$ and $h$, for $E_c=0.05\l|\mu\r|$, $\lambda^{\mathrm{inter}}=0.2$, and $\omega_c=0.2\l|\mu\r|$.}
    \label{fig:Tc_vshgamma}
\end{figure}

Having determined the value of $h$ which produces a maximum of the overlap in the different cases, with our choice of parameters we calculate numerically the critical temperature of the interband domains as a function of $\gamma$ and $h$. The numerical results are plotted in \cref{fig:Tc_vshgamma} together with the theoretical prediction given by \cref{eq:hpos1d2u,eq:hpos1u2d}. 

We observe that for a good range of $\gamma$ values ($0.95\lesssim\gamma\lesssim1.05$ and $1.15\lesssim\gamma\lesssim1.3$) the numerical results follow the analytical results and superconductivity comes from case (i) or case (ii), showing peaks in $T_c$ for non-zero values of $h$. Instead, for $1.06<\gamma<1.14$, the superconducting domains are centered in $h=0$, away from the analytical prediction, except for $\gamma\simeq1.1$. Therefore, for this choice of parameters, and this range of $\gamma$ values, the system favors superconductivity in absence of spin-splitting. Considering instead those points which do not exhibit superconductivity, for $\gamma<0.95$ and $\gamma>1.3$, we can state that the overlap between the bands is not large enough for the chosen value of the interband superconducting coupling constant. However, choosing a higher coupling would yield superconductivity for values of $\gamma$ smaller than 0.95 and higher than 1.3.


Observing \cref{fig:Tc_vshgamma}, we can see that peaks in $T_c$ can be found for rather high values of the exchange field, up to $h\simeq1.6E_c$. Therefore, when the system has both intraband and interband superconducting pairing, it can present two disconnected superconducting domains, one for low magnetic field due to the intraband pairing, and one for high magnetic field coming exclusively from interband pairing.

Finally, it is worth noting that the case of two electron-like bands would yield qualitatively similar results. The difference would be in the value of the band mass ratio $\gamma$ at which the band overlap occurs. The case where there is a coupling between an electron-like and a hole-like band is more complicated and would require a separate study.


\section{Minimal Models for two-dimensional \texorpdfstring{M\lowercase{g}B\textsubscript{2} and B\lowercase{a}\textsubscript{0.6}K\textsubscript{0.4}F\lowercase{e}\textsubscript{2}A\lowercase{s}\textsubscript{2}}{MgB2 and BaKFe2As2}}

\begin{figure}
    \centering
    \includegraphics[width=\columnwidth]{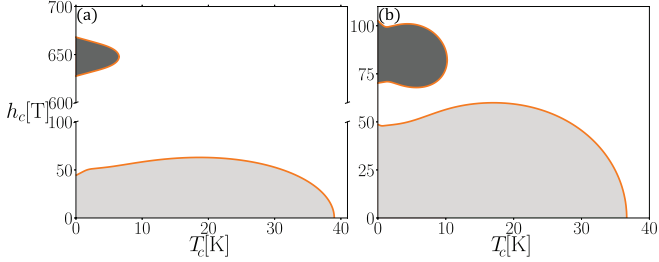}
    \caption{Superconducting critical curves from a minimal model for (a) two-dimensional MgB\textsubscript{2} and (b) Ba\textsubscript{0.6}K\textsubscript{0.4}Fe\textsubscript{2}As\textsubscript{2}, with the highlighted high (dark gray) and low (light) field superconducting domains.}
    \label{fig:MgB2_BaKFe2As2_SC}
\end{figure}

The simple model presented in this work, for an ideal case, can be realized in a 2D multiband superconductor with the application of an in-plane external magnetic field. On the material side many superconductors have been shown to have multiple hole-like bands close to the Fermi energy at the $\Gamma$ point. For instance, band structure calculations in few monolayers (MLs) MgB\textsubscript{2} in Ref.~\cite{Bekaert2017} have shown that each ML contributes to the band structure with a pair of $\sigma$ bands, $\sigma_1$ and $\sigma_2$ having different effective masses and chemical potential $\mu\simeq500\si{meV}$, with each pair of bands separated by $E_c\simeq37.5\si{meV}$. Therefore, we consider a minimal model for 2ML two-dimensional MgB\textsubscript{2}, taking only a pair of $\sigma$ bands separated by $37.5\si{meV}$. In this case we retain the superconducting coupling constants representing scattering processes connecting interband to intraband Cooper pairs and vice versa, which were previously neglected. This means that in this case the intraband and interband order parameters are coupled, the resulting set of linearized gap equations is given in \cref{appendix:interintra_gapeq}. We reduce the indices of the $\lambda_{\alpha\alpha',\beta\beta'}$ coupling constants in the following way:

\begin{equation*}
\alpha\alpha'\equiv
\begin{cases}
    &\alpha\quad\mathrm{if}\quad\alpha=\alpha',\\
    &3\quad\mathrm{if}\quad\alpha\neq\alpha'.
\end{cases}
\end{equation*}

\noindent Therefore, the index 3 represents the interband pairing channel. Following Ref.\cite{Vargas-Paredes2020} we consider the following superconducting coupling matrix:

\be
    \hat{\lambda}=\begin{pmatrix} 0.275 & 0.032 & \lambda_{13} \\ 0.032 & 0.274 & \lambda_{23} \\ \lambda_{31} & \lambda_{32} & \lambda_{33} \end{pmatrix},
\ee

\noindent where the upper $2\times2$ block represent the conventional superconducting coupling for a two band system, the terms $\lambda_{\alpha3}$ and $\lambda_{3\alpha}$, $\alpha=1,2$, represent interband Cooper pairs scattering to intraband Cooper pairs and vice versa. Finally $\lambda_{33}$ represents purely interband scattering processes, and corresponds to the constant $\lambda^{\mathrm{inter}}$. To obtain the correct value of the critical temperature at zero field ($T_c\approx39\si{K}$), we set $\lambda_{\alpha3}=\lambda_{3\alpha}=0.045$ and $\lambda_{33}=0.2$. In \cref{fig:MgB2_BaKFe2As2_SC}(a) we report the superconducting domain delimited by the $h_c$-$T_c$ critical curves. As in the previous section we find two disconnected superconducting domains for low and high magnetic field, with the high magnetic field one centered at a value $B=\mu_BE_c\approx650\si{T}$. 

Also, among FeBS, many materials in the iron-pnictides family, like LaFeAsO\textsubscript{$1-x$}F\textsubscript{$x$} \cite{Xu2008,Mazin2008,Haule2008} and Ba\textsubscript{$1-x$}K\textsubscript{$x$}Fe\textsubscript{2}As\textsubscript{2} \cite{Ding2011}, exhibit multiple hole-like bands at the $\Gamma$ points, mainly originating from the 3-\textit{d} Fe orbitals. Again, we consider a minimal model for optimally doped Ba\textsubscript{0.6}K\textsubscript{0.4}Fe\textsubscript{2}As\textsubscript{2}, with the two hole-like $\alpha$ and $\beta$ bands at the $\Gamma$ point, separated by $E_c=5\si{meV}$ \cite{Ding2011}. Following Ref.\cite{Vargas-Paredes2020} we take the superconducting coupling matrix to be:

\be
    \hat{\lambda}=\begin{pmatrix} 0.51 & -0.005 & \lambda_{13} \\ -0.0025 & 0.39 & \lambda_{23} \\ \lambda_{31} & \lambda_{32} & \lambda_{33} \end{pmatrix},
\ee

\noindent where the elements 12 and 21 are set to negative values as required by the $s_\pm$ superconducting phase typical of FeBS. However, we note that the critical temperature and field of the system are not sensitive to this negative sign since the intraband gaps are effectively decoupled from the interband gap for the parameters used. Once more, we set $\lambda_{\alpha3}=\lambda_{3\alpha}=0.05$, $\alpha=1,2$, and $\lambda_{33}=0.25$ to obtain the right critical temperature at zero field ($T_c\approx36\si{K}$) and report the results in \cref{fig:MgB2_BaKFe2As2_SC}(b). Again we note two disconnected superconducting domains with the high field one centered around $h_c\approx90\si{T}$.

\section{Conclusions}

In summary, we have considered a superconducting system with two relevant low-energy bands in the presence of a spin-splitting field. We have shown that the inclusion of interband superconducting pairing processes, \ie allowing for the presence of Cooper pairs with the two electrons coming from two different bands, can result in reentrant superconducting domains centered at a high value of the spin-splitting field. These domains are exclusively due to the interband pairing and generally do not overlap with the superconducting domains originating from intraband pairing, which are always centered at zero spin-splitting field.

\begin{acknowledgments}
Computations have been performed on the SAGA supercomputer provided by UNINETT Sigma2 - the National Infrastructure for High Performance Computing and Data Storage in Norway. We acknowledge funding via the “Outstanding Academic Fellows” programme at NTNU, the Research Council of Norway Grant No. 302315, as well as through its Centres of Excellence funding scheme, Project No. 262633, “QuSpin.” Nordita is supported in part by NordForsk.
\end{acknowledgments}

\appendix

\section{Green's function of the system.}
\label{appendix:Green}
The Green's function of the system is obtained by inverting \cref{eq:invG} in the main text:

\be
	\bm{\mathcal{G}}_0(\bm{k},\oMn)=\frac{1}{\mathrm{det}\bm{\mathcal{G}}_0^{-1}}\begin{pmatrix} \hat{\mathcal{A}}^{11}(\bm{k},\tilde{\omega}_n) & \hat{\mathcal{A}}^{12}(\bm{k},\tilde{\omega}_n) \\ \hat{\mathcal{A}}^{21}(\bm{k},\tilde{\omega}_n) & \hat{\mathcal{A}}^{22}(\bm{k},\tilde{\omega}_n) \end{pmatrix},
\ee

\noindent where $\mathrm{det}\bm{\mathcal{G}}_0^{-1}=\l(\tilde{\omega}_n^2+E_+^2\r)\l(\tilde{\omega}_n^2+E_-^2\r)$ with $\tilde{\omega}_n=\oMn-ih$. The two branches have the following expression:

\begin{widetext}
\be
E_\pm^2\!=\!\frac{1}{2}\!\l\{\!\tilde{E}_1^2\!+\!\tilde{E}_2^2\!+\!2\l|\Delta_{12}\r|^2\pm\!\l[\l(\tilde{E}_1^2\!+\!\tilde{E}_2^2\!+\!2\l|\Delta_{12}\r|^2\r)^2\!-\!4\l(\tilde{E}_1^2\tilde{E}_2^2\!+\!\l|\Delta_{12}\r|^4+\!2\tilde{\xi_1}\tilde{\xi_2}\l|\Delta_{12}\r|^2\!-\!2\mathrm{Re}\l\{\Delta_{11}\Delta_{22}{\Delta_{12}^*}^2\r\}\r)\r]^{\frac{1}{2}}\r\}.
\label{eq:bcs_qp}
\ee
\end{widetext}

\noindent The the components of the $2\times2$ matrices $\hat{\mathcal{A}}^{\alpha\beta}(\bm{k},\oMn)$ are:

\bse\begin{align}
	\l[\hat{\mathcal{A}}^{11}(\bm{k},\tilde{\omega}_n)\r]_{11}=&-\l(i\tilde{\omega}_n+\tilde{\xi_1}\r)\!\l(\tilde{\omega}_n^2\!+\!\tilde{\xi_2}^2\!+\!\l|\Delta_{22}\r|^2\!\r) \notag \\
        &-\!\l|\Delta_{12}\r|^2\l(i\tilde{\omega}_n+\tilde{\xi_2}\r),\\
	\l[\hat{\mathcal{A}}^{11}(\bm{k},\tilde{\omega}_n)\r]_{12}=&\Delta_{11}\l(\tilde{\omega}_n^2+\tilde{\xi_2}^2+\l|\Delta_{22}\r|^2\!\r)-\Delta_{12}^2\Delta_{22}^*,\\
	\l[\hat{\mathcal{A}}^{12}(\bm{k},\tilde{\omega}_n)\r]_{11}=&\l(i\tilde{\omega}_n+\tilde{\xi_1}\r)\Delta_{12}\Delta_{22}^* \notag\\
        &+\Delta_{11}\Delta_{12}^*\l(i\tilde{\omega}_n+\tilde{\xi_2}\r),\\
	\l[\hat{\mathcal{A}}^{12}(\bm{k},\tilde{\omega}_n)\r]_{12}=&\!-\!\l[\l(i\tilde{\omega}_n\!+\!\tilde{\xi_1}\r)\!\l(i\tilde{\omega}_n\!-\!\tilde{\xi_2}\r)\!-\!\l|\Delta_{12}\r|^2\r]\Delta_{12} \notag \\
        &-\Delta_{11}\Delta_{12}^*\Delta_{22},\\
	\l[\hat{\mathcal{A}}^{21}(\bm{k},\tilde{\omega}_n)\r]_{11}=&\l[\hat{\mathcal{A}}^{12}(\bm{k},-\tilde{\omega}_n^*)\r]_{11}^*,\\
	\l[\hat{\mathcal{A}}^{21}(\bm{k},\tilde{\omega}_n)\r]_{12}=&\l[\hat{\mathcal{A}}^{12}(\bm{k},-\tilde{\omega}_n)\r]_{12},\\
	\l[\hat{\mathcal{A}}^{\alpha\beta}(\bm{k},\tilde{\omega}_n)\r]_{22}=&-\l[\hat{\mathcal{A}}^{\alpha\beta}(\bm{k},-\tilde{\omega}_n)\r]_{11},\\
	\l[\hat{\mathcal{A}}^{\alpha\beta}(\bm{k},\tilde{\omega}_n)\r]_{21}=&\l[\hat{\mathcal{A}}^{\alpha\beta}(\bm{k},\tilde{\omega}_n^*)\r]_{12}^*.
\end{align}\ese

\noindent where $\tilde{E}_\alpha=\sqrt{\tilde{\xi}_\alpha^2+\l|\Delta_{\alpha\alpha}\r|^2},\,\,\alpha=1,2$, with $\tilde{\xi_1}=\x{1}-E_c,\,\tilde{\xi_2}=\x{2}+E_c$.

The elements of $\hat{\mathcal{A}}^{22}(\bm{k},\oMn)$ are obtained by exchanging the indices 1 and 2 in the expressions for the elements of $\hat{\mathcal{A}}^{11}(\bm{k},\oMn)$.
The inter- and intra-band normal and anomalous Green's functions are given by:

\bse\begin{align}
	G^{\alpha\beta}(\bm{k},\tilde{\omega}_n)&=\frac{\l[\hat{\mathcal{A}}^{\alpha\beta}(\bm{k},\tilde{\omega}_n)\r]_{11}}{\l(\tilde{\omega}_n^2+E_+^2\r)\l(\tilde{\omega}_n^2+E_-^2\r)},\\
	F^{\alpha\beta}(\bm{k},\tilde{\omega}_n)&=\frac{\l[\hat{\mathcal{A}}^{\alpha\beta}(\bm{k},\tilde{\omega}_n)\r]_{12}}{\l(\tilde{\omega}_n^2+E_+^2\r)\l(\tilde{\omega}_n^2+E_-^2\r)}.
\end{align}\ese

\section{Gap equation for the interband order parameter.}
\label{appendix:inter_gapeq}

We derive the gap equation in the case of purely interband coupling, \ie Cooper pairs formed exclusively by electrons in different bands. To do so we consider the following form of the coupling matrix:

\be
g_{\alpha\alpha'\beta\beta'}=\begin{cases} g^{\mathrm{inter}}, \quad \mathrm{if}\quad \alpha\neq\alpha',\,\beta\neq\beta'\\ 0, \quad \mathrm{otherwise}. \end{cases}
\ee

\noindent Inserting this in \cref{eq:gapEq} we obtain the following gap equation:

\be
    \Delta_{12}=\frac{g^{\mathrm{inter}} T}{V}\sum_{\bm{k}\oMn}\l[F^{12}(\bm{k},\oMn-ih)+F^{21}(\bm{k},\oMn-ih)\r].
    \label{eq:inter_gapeq}
\ee

\noindent Setting $\Delta_{11}=\Delta_{22}=0$ in \cref{eq:bcs_qp} we get:

\be
    E_\pm=\frac{\tilde{\xi}_1-\tilde{\xi}_2}{2}\pm E_{12},
\ee

\noindent with $E_{12}=\sqrt{(\tilde{\xi}_1+\tilde{\xi}_2)^2/4+\l|\Delta_{12}\r|^2}$. Using the relation, defined in the main text, between the two bands $\x{2}=\gamma\x{1}+(\gamma-1)\mu$, we can write:

\bse\begin{align}
    E_\pm&=\frac{(\x{1}+\mu)(1-\gamma)}{2}-E_c\pm E_{12},\\
    E_{12}&=\sqrt{\l(\frac{\x{1}(1+\gamma)-\mu(1-\gamma)}{2}\r)^2+\l|\Delta_{12}\r|^2}.
\end{align}\ese

The expression of the anomalous Green's function is:

\be\begin{split}
    F^{12}&(\bm{k},\oMn-ih)=\\&-\frac{\l(i\tilde{\omega}_n+\tilde{\xi_1}+h\r)\l(i\tilde{\omega}_n-\tilde{\xi_2}+h\r)-\l|\Delta_{12}\r|^2}{\l[(\oMn-ih)^2+E_+^2\r]\l[(\oMn-ih)^2+E_-^2\r]}\Delta_{12}.
\end{split}\ee

After summing over the Matsubara frequency the gap equation \cref{eq:inter_gapeq} takes the following form:

\begin{widetext}
\be\begin{split}
    \frac{1}{g^{\mathrm{inter}}}=&-\frac{1}{V}\sum_{\bm{k}}\frac{1}{2E_{12}}\l[n_F(E_+-h)-n_F(-E_+-h)-n_F(E_--h)+n_F(-E_--h)\r]\\
    =&\frac{1}{V}\sum_{\bm{k}}\sum_{s=\pm}\frac{s}{4E_{12}}\l\{\tanh\l[\frac{\beta}{2}\l(E_s+h\r)\r]+\tanh\l[\frac{\beta}{2}\l(E_s-h\r)\r]\r\},
\end{split}\ee
\end{widetext}

\noindent where $n_F(\varepsilon)=(e^{\beta\varepsilon}+1)^{-1}$ is the Fermi function and we have used:

\begin{equation*}\begin{split}
    n_F&(E_\pm-h)-n_F(-E_\pm-h)\\&=-\frac{\sinh\beta E_\pm}{\cosh\beta E_\pm+\cosh\beta h}\\&=-\frac{1}{2}\l\{\tanh\l[\frac{\beta}{2}\l(E_\pm+h\r)\r]+\tanh\l[\frac{\beta}{2}\l(E_\pm-h\r)\r]\r\}.
\end{split}\end{equation*}

\noindent We now switch from the summation over momenta to integral over the energy: $V^{-1}\sum_{\bm{k}}(\cdot)=\int d\xi_1N_1(\xi_1)(\cdot)$, where $N_1(\xi_1)$ is the density of state of band 1. We then approximate the density of state with its value at the Fermi level $N_1(\xi_1)\simeq N_1(0)$ and, defining the dimensionless superconducting coupling constant as done in the main text, we obtain the final expression for the interband gap equation:

\be\begin{split}
    \frac{1}{\lambda^{\mathrm{inter}}}&=\int_{-\omega_c}^{\omega_c}d\xi\sum_{s=\pm}\frac{s}{4E_{12}(\xi)}\\
    \times&\l\{\tanh\l[\frac{\beta}{2}\l(E_s(\xi)+h\r)\r]+\tanh\l[\frac{\beta}{2}\l(E_s(\xi)-h\r)\r]\r\}.
    \label{eq:Sinter_gapeg}
\end{split}\ee

\noindent To obtain critical temperature and critical field, we linearize \cref{eq:Sinter_gapeg} with respect to $\Delta_{12}$. The equation then takes the following simple form:

\begin{widetext}
\be
    \frac{1}{\lambda^{\mathrm{inter}}}=\int_{-\omega_c}^{\omega_c}\frac{d\xi}{2(\tilde{\xi}_1+\tilde{\xi}_2)}\sum_{\alpha=1,2}\l\{\tanh\l[\frac{\beta_c}{2}\l(\tilde{\xi}_\alpha+h_c\r)\r]+\tanh\l[\frac{\beta_c}{2}\l(\tilde{\xi}_\alpha-h_c\r)\r]\r\}.
    \label{eq:Slingapinter}
\ee
\end{widetext}

\noindent We note that by setting $E_c=0$ and $\gamma=1$, we get $\tilde{\xi}_1=\tilde{\xi}_2=\xi$, the system reduces to a single band superconductor. Therefore, \cref{eq:Slingapinter} takes the usual form for a single band spin-split superconductor:

\be
    \frac{1}{\lambda}\!=\!\int_{-\omega_c}^{\omega_c}\frac{d\xi}{2\xi}\l\{\!\tanh\!\l[\!\frac{\beta_c}{2}\!\l(\xi+h_c\r)\!\r]\!+\!\tanh\!\l[\!\frac{\beta_c}{2}\!\l(\xi-h_c\r)\!\r]\!\r\}.
    \label{eq:Slingap_1b}
\ee

\section{Gap equation for the intraband order parameters.}
\label{appendix:intra_gapeq}

In order to consider purely intraband coupling we take the following form of the coupling matrix:

\be
g_{\alpha\alpha'\beta\beta'}=\begin{cases} g_{\alpha\beta}^{\mathrm{intra}}, \quad \mathrm{if}\quad \alpha=\alpha',\,\beta=\beta'\\ 0, \quad \mathrm{otherwise} \end{cases}
\ee

\noindent Using this expression in \cref{eq:gapEq} we get the following coupled gap equations:

\bse\begin{align}
    \Delta_{11}=&\frac{g_{11}^{\mathrm{intra}}T}{V}\sum_{\bm{k}\oMn}F^{11}(\bm{k},\oMn-ih) \notag \\
        &+\frac{g_{12}^{\mathrm{intra}}T}{V}\sum_{\bm{k}\oMn}F^{22}(\bm{k},\oMn-ih),\\
    \Delta_{22}=&\frac{g_{21}^{\mathrm{intra}}T}{V}\sum_{\bm{k}\oMn}F^{11}(\bm{k},\oMn-ih) \notag \\
        &+\frac{g_{22}^{\mathrm{intra}}T}{V}\sum_{\bm{k}\oMn}F^{22}(\bm{k},\oMn-ih).
\end{align}\ese

\noindent The two branches of the BCS quasiparticle spectrum are simply $E_+=E_{11}$ and $E_-=E_{22}$ and we get the following expression for the anomalous Green's function:

\be
    F^{\alpha\alpha}(\bm{k},\oMn-ih)=\frac{\Delta_{\alpha\alpha}}{\l(i\oMn+E_{\alpha\alpha}+h\r)\l(i\oMn-E_{\alpha\alpha}+h\r)}.
\ee

\noindent Following the same steps as in the previous section we finally get the following:

\be\begin{split}
    \Delta_{\alpha\alpha}=\sum_{\beta=1,2}&\lambda_{\alpha\beta}\Delta_{\beta\beta}\int_{-\oMn}^{\oMn}\frac{d\xi_{\beta}}{4E_{\beta\beta}}\\
    &\times\l(\tanh\frac{E_{\beta\beta}+h}{2T}+\tanh\frac{E_{\beta\beta}-h}{2T}\r).
    \label{eq:Sintra_gapeg}
\end{split}\ee

\noindent Linearizing \cref{eq:Sintra_gapeg} with respect to the order parameters, again allows to obtain the superconducting critical temperature and critical field. We set $\Delta_{\alpha\alpha}=\epsilon\delta_\alpha$, where $\epsilon(T=T_c)=0$, and obtain the following system:

\be
    \begin{pmatrix} \delta_1 \\ \delta_2 \end{pmatrix}\begin{pmatrix} \lambda_{11}I_1(T_c,h_c)-1 & \lambda_{12}I_2(T_c,h_c) \\ \lambda_{21}I_1(T_c,h_c) & \lambda_{22}I_2(T_c,h_c)-1 \end{pmatrix}=0,
    \label{eq:Slingapintra}
\ee

\noindent where

\be\begin{split}
    I_\alpha&(T,h)=\\
    \int_{-\omega_c}^{\omega_c}&\frac{d\xi}{4\tilde{\xi}_\alpha}\l\{\tanh\l[\frac{\beta}{2}\l(\tilde{\xi}_\alpha+h\r)\r]+\tanh\l[\frac{\beta}{2}\l(\tilde{\xi}_\alpha-h\r)\r]\r\},
    \label{eq:Sintegrals_intra}
\end{split}\ee

\noindent with $\alpha=1,2$. The critical parameters are found by setting to zero the determinant of the matrix in \cref{eq:Slingapintra}.

\section{Linearized gap equation for coupled interband and intraband order parameters}
\label{appendix:interintra_gapeq}

When including the superconducting coupling constants representing scattering processes connecting interband to intraband Cooper pairs, the gap equations for the interband and intraband order parameters become coupled. In this case the linearized gap equation yields the following system:

\begin{widetext}
\be
    \begin{pmatrix} \delta_1 \\ \delta_2 \\ \delta_3 \end{pmatrix}\begin{pmatrix} \lambda_{11}I_1(T_c,h_c)-1 & \lambda_{12}I_2(T_c,h_c) & \lambda_{13}I_3(T_c,h_c)\\ \lambda_{21}I_1(T_c,h_c) & \lambda_{22}I_2(T_c,h_c)-1 & \lambda_{23}I_3(T_c,h_c) \\ \lambda_{31}I_1(T_c,h_c) & \lambda_{32}I_2(T_c,h_c) & \lambda_{33}I_3(T_c,h_c)-1 \end{pmatrix}=0,
    \label{eq:Slingapcoupled}
\ee
\end{widetext}

\noindent where the constants $\lambda_{\alpha3}$ and $\lambda_{3\alpha}$, with $\alpha=1,2$, represent the scattering processes connecting interband to intraband Cooper pairs, and $\lambda_{33}$ corresponds to $\lambda^{\mathrm{inter}}$ in the main text. The terms $I_\alpha$, $\alpha=1,2$ are given in \cref{eq:Sintegrals_intra}, while $I_3$ corresponds to the energy integral in \cref{eq:Slingapinter}. Again, the critical parameters are found by setting to zero the determinant of the matrix in \cref{eq:Slingapcoupled}.

\section{Critical Temperature and Critical Field.}
\label{appendix:hc(T)}

We presented the linearized equations \cref{eq:Slingapinter,eq:Slingapintra} allowing to obtain the curve $h_c(T)$, for the interband and intraband domains, respectively. A remark here is needed, since as is clear from \cref{fig:comb_hc_Ec0.05} in the main text, in the interband curves, to each temperature (except the maximum one), correspond two solutions for $h$. This is a problem for the numerical solver. To address this, first we find the value of the critical field corresponding to the maximum temperature of the superconducting curve. This can be done through analytical considerations as explained in the main text (see \cref{eq:hpos1d2u,eq:hpos1u2d}). Having obtained this value we then find separately a solution for the critical field higher and lower than this value, as a function of the temperature.


%

\end{document}